\documentclass[aps, prl,twocolumn,groupedaddress]{revtex4-2}
\usepackage{CJK}
\usepackage{color}
\usepackage{graphicx}
\usepackage{dcolumn}
\usepackage{subfigure}
\usepackage{graphics}
\usepackage{amsmath}
\usepackage{amsfonts}
\usepackage{amssymb}
\usepackage{epsf}
\usepackage{bm}

\begin{document}
\begin{CJK*}{GB}{ } % Use default fonts from CJK (see below)

\title{Quantifying the impact of state mixing on the Rydberg excitation blockade}
\author{Milo Eder$^{1}$, Andrew Lesak$^{1}$, Abigail Plone$^{1}$, Tomohisa Yoda$^{1}$, Michael Highman$^{2\dag}$, and Aaron Reinhard$^{1*}$}

\address{
$^{1}$Department of Physics, Kenyon College, 201 North College Rd., Gambier, Ohio 43022, USA \\
$^{2}$Department of Physics, Otterbein University, 1 South Grove St., Westerville, OH 43081, USA \\
$^*$Corresponding author: reinhard1@kenyon.edu\\
$^{\dag}$Present address: Department of Physics, University of Illinois at Urbana-Champaign, Urbana, Illinois 61801-3080, USA
}

\date{\today}
\begin{abstract}
The Rydberg excitation blockade has been at the heart of an impressive array of recent achievements; however, state-mixing interactions can compromise its efficiency.  When ultracold atoms are excited to Rydberg states near F\"orster resonance, up to $\sim 50\%$ of atoms can be found in dipole coupled product states within tens of ns after excitation.  There has been disagreement in the literature regarding the mechanism by which this mixing occurs.  We use state-selective field ionization spectroscopy to measure, on a shot-by-shot basis, the distribution of Rydberg states populated during narrowband laser excitation.  Our method allows us to both determine the number of additional Rydberg excitations added by each mixing event, and to quantify the extent to which state mixing ``breaks'' the blockade. For excitation of ultracold rubidium atoms to $nD_{5/2}$ states, we find that the mixing is consistent with a three-body process, except near exact F\"orster resonance.
\end{abstract}

\maketitle		
\end{CJK*}

The Rydberg excitation blockade is a coherent effect that results from electrostatic interactions among atoms in high-lying states~\cite{Jaksch,Tong,Cubel}.  If an ensemble of $N$ ultracold atoms is excited to a state with $k$ shared Rydberg excitations, interactions cause the energy spectrum of the collective states $\vert N, k \rangle$ to be unequally spaced.  The blockade occurs if the system is driven by a narrowband laser resonant with the $\vert N, 0 \rangle \rightarrow \vert N, 1 \rangle$ transition.  Excitation to $k \geq 2$ states is suppressed because these states are shifted by more than $h \times \delta_L$, where $\delta_L$ is the laser linewidth.  If the blockade condition is met throughout the entire sample, exactly one Rydberg excitation is created~\cite{Urban,Gaetan,Wilk,Dudin,Ebert,Labhun}.  However, larger samples break up into ``single-excitation domains,'' inside of which all atoms coherently share one excitation~\cite{Cubel,Reinhard_DR,Reinhard3,Schwarzkopf,Schauss}.  The diameter of these domains represents the correlation distance between the measured positions of the excitations.

The Rydberg blockade has been employed in a wide range of applications in the last decade.  These include the implementation of few-body entanglement~\cite{Urban,Gaetan,Dudin,Li,Barredo}, strongly correlated systems~\cite{Schwarzkopf,Schauss,Urvoy}, efficient single photon sources~\cite{Dudin2}, single photon switches and transistors~\cite{Baur,Tiarks}, optical nonlinearities~\cite{Pritchard,Hofmann,Maxwell}, direct measurement of dipole-dipole interaction energies~\cite{Altiere,Beguin}, and coherent evolution in room temperature vapors~\cite{Urvoy,Baluktsian}.  However, many exciting applications of the blockade have been to neutral atom quantum information and quantum simulation~\cite{Jaksch,Lukin,Saffman_Rev}. Tremendous progress has been made in using site addressable single atoms in lattices or optical tweezer arrays~\cite{Isenhower,Wang,Wang2,Labhun,Weitenberg,Bernien,Lienhard,deLesluec}; however, larger clouds have also been proposed to implement quantum gates \cite{Jaksch,Lukin}.

For the blockade to be used optimally, processes that reduce its efficiency must be carefully quantified.  For example, in Refs.~\cite{Walker,Walker2} the authors showed that, for certain quantum numbers, there exist linear combinations of pair states with different $m_j$ that cause the dipole matrix element with a nearby state to vanish.  These so-called ``F\"orster zero'' states lead to reduced pair-state energy shift.  Another example can be found in Refs.~\cite{Labhun,deLesluec}, where the authors showed that certain combinations of electric and magnetic fields cause the pair potentials to cross zero at small separation, leading to resonant excitation of pairs within a domain.  A third example is the so-called ``antiblockade'' where a detuned excitation laser can lead to enhanced excitation of atom pairs at a given separation~\cite{Amthor,Urvoy}. In the present work, we characterize the negative impact of state-mixing interactions on the blockade near F\"orster resonance.  This is important to quantify, since these interactions could cause decoherence in experiments that do not have state-selective Rydberg atom detection~\cite{Labhun,deLesluec}.

We consider the interaction channel $2 \times nD_{5/2} \rightarrow (n-2)F_{7/2} + (n+2)P_{3/2}$ in rubidium, which becomes nearly resonant at $n=43$~\cite{Reinhard}.  This leads to an enhancement of the interaction energies, ostensibly making $43D_{5/2}$ states a good candidate for applications of the blockade.  However, in Refs.~\cite{Reinhard2,Younge} it was shown that up to $\sim 50\%$ of the population could be detected in the $41F$ and $45P$ product states within 100ns of excitation to $43D_{5/2}$ in zero electric field.  Since each product state is detuned from the excitation laser by $\sim \pm 62.5$~GHz, they can only be created together.  This, by definition, breaks the blockade.  The authors concluded that sums over pairwise potentials could not account for the large magnitude of mixing, but a ``complete basis many body theory'' was needed.  Later that year, a model for state mixing was proposed which featured optical pumping to a ``dark state'' in a manifold of three-particle states~\cite{Pohl}.  The dark state is characterized by one atom each in the $nD_{5/2}$, $(n-2)F_{7/2}$, and $(n+2)P_{3/2}$ states, suggesting that state mixing is a three-body process.  However, several years later, Kondo \textit{et al} performed an experiment similar to the one in Refs.~\cite{Reinhard2,Younge}.  They showed that the number of $(n+2)P_{3/2}$ Rydberg excitations created near resonance scales with the total Rydberg population squared.  They concluded that state mixing is a two-body process, although they proposed no mechanism to explain the large rates~\cite{Kondo}.  To date, this discrepancy has not been resolved.

In Refs.~\cite{Reinhard2,Younge,Kondo} the fraction of atoms in product states just after excitation was found from an average over many experimental cycles.  In our experiment, we observe statemixing on a shot-by-shot basis and record the number of additional Rydberg excitations added each time a single mixing event occurs.  This allows us to observe directly whether state mixing near F\"orster resonance is a two-body or three-body process. This is an important question, since it is unknown whether two qubit gates are immune.  In contrast to Ref.~\cite{Kondo}, we find that state mixing is consistent with a three-body process~\cite{Pohl}, except near exact F\"orster resonance where an unfavorable pair potential can lead to resonant excitation of pairs.

Our method has a further advantage.  While it is understood theoretically that state mixing reduces the efficiency of the blockade~\cite{Younge,Pohl}, the extent of its negative impact has not been quantified. Large amounts of mixing, which adds unwanted excitations, have always accompanied an improved blockade near F\"orster resonance~\cite{Reinhard,Younge,Reinhard3}. The blockade-enhancing effect of larger interaction energies has been entangled with the blockade-ruining effect of state mixing.  By observing the mixing on an event-by-event basis, we can disentangle these two effects and quantify, for the first time, the extent to which state-mixing interactions reduce the blockade efficiency.  We find that the number of extra excitations added by state mixing is smallest on resonance, where the probability to have a state-mixing event is highest.  This lends support to the notion that ``excitation domains'' remain a useful concept, even in the presence of significant mixing.

\begin{figure}[htp]
\centerline{ \scalebox{0.56} {\includegraphics{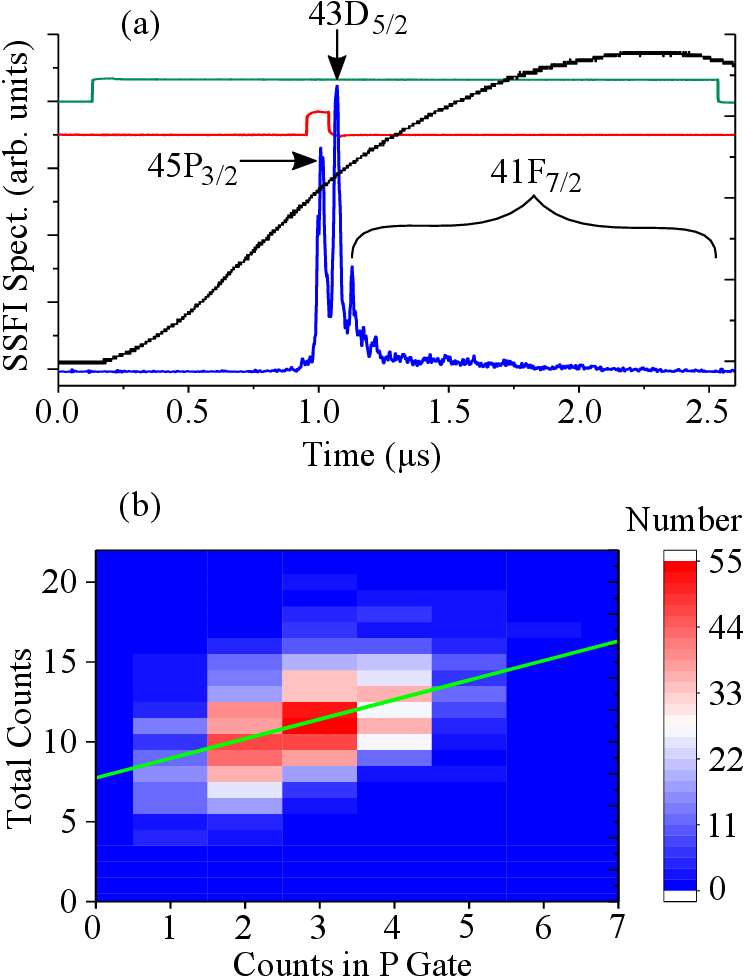}}} \caption{ \label{Fig1}  (a) Average SSFI spectrum for excitation to $43D_{5/2}$, the T (green) and P (red) gates which have been offset and scaled for clarity, and the SSFI ramp.   (b)  An example ``sorted'' graph: the total number of excitations ($N_T$) as a function of the number in $(n+2)P$ ($N_P$).  The false color shows how many of each $\{N_P, N_T\}$ were detected.  The green line is a fit, from which we extract the slope and intercept.}
\end{figure}

We cool and trap $^{85}$Rb atoms in a vapor cell magneto-optical trap (MOT).  We focus an 8.9~W, $1064$~nm laser beam through the cloud to form an optical dipole trap with peak density $1.2 \times 10^{11}$~cm$^{-3}$ and RMS widths of 7.5~$\mu$m and 220~$\mu$m along the horizontal and vertical directions.  The ground state atom density at the time of excitation is controlled by turning off the dipole trap beam and waiting a variable time (0 to 180~$\mu$s) for the atom cloud to expand~\cite{Younge}.  We then apply two coincident, 200ns excitation pulses to drive the $5S_{1/2} \rightarrow 6P_{3/2} \rightarrow nD_{5/2}$ transitions.  The lower transition beam, with $\sigma +$ polarization and wavelength 420.298~nm, is focused through the short axis of the cloud to a waist of 40~$\mu$m with Rabi frequency 17~MHz.  The upper transition beam, with $\pi$ polarization and wavelength $\approx 1017$~nm, is focused to a waist of 35~$\mu$m with Rabi frequency 19~MHz, perpendicular to both the atom cloud's long axis and the lower transition beam.  The upper transition laser is locked to a pressure-tunable Fabry P\'erot cavity, which allows us to frequency stabilize and tune its output~\cite{Orr}.  The beams are detuned from the intermediate $6P_{3/2}$ level by 50~MHz, leading to a two-photon Rabi frequency of $3$~MHz.  Our lasers have linewidth 1.5 MHz, yielding a total excitation linewidth of 3.5~MHz.  We perform state-selective field ionization (SSFI) spectroscopy 50~ns after excitation by applying a high voltage ramp across a pair of electrodes above and below the atom cloud.  The Rydberg atoms are ionized and the electrons are accelerated vertically to a dual stage microchannel plate detector.  We count pulses from single electrons using a gated pulse counter.

Our measurement proceeds as follows.  We first average all pulses from 750 experimental sequences to obtain an averaged SSFI spectrum, shown in Fig.~\ref{Fig1}a for excitation to $43D_{5/2}$.  The central peak corresponds to atoms in this state, while the left peak corresponds to atoms in $45P_{3/2}$.  We use the averaged SSFI spectrum to define two counting gates.  The ``P Gate'' is set to count only atoms in $(n+2)P_{3/2}$, while the ``T Gate'' is set to count all Rydberg atoms.  We then record, for each of 1001 experimental sequences, the total number of Rydberg atoms created, as well as the number in the P gate.  We plot the total number of excitations as a function of the number in the $(n+2)P_{3/2}$ state, as shown in Fig.~\ref{Fig1}b.  We then fit this ``sorted graph'' to a line.  The slope tells us how many extra Rydberg excitations are added each time an atom is mixed into the $(n+2)P_{3/2}$ state. This allows us to determine whether state mixing is dominated by a two-body or three-body process.  The intercept tells us how many excitations \emph{would} be created if there were no state mixing.   The difference between the number of excitations actually created and the y-intercept tells us how many additional excitations state mixing adds to the system.  This quantifies the extent to which state mixing breaks the blockade.

\onecolumngrid
\begin{center}
\begin{figure}[htp]
\centerline{ \scalebox{0.365} {\includegraphics{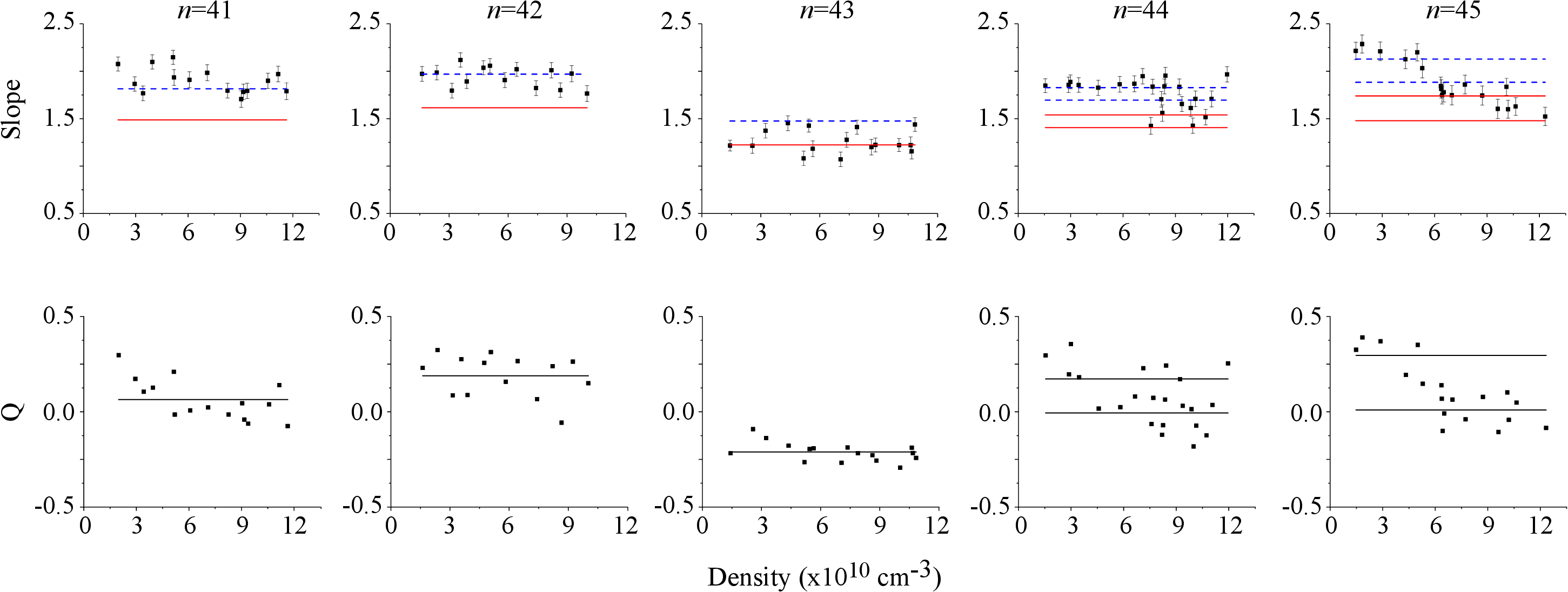}}} \caption{ \label{Fig2} Top panels: Slopes of the fits to the sorted data as a function of ground state atom density.  The error bars are the uncertainties in the fitted slopes.  Horizontal lines are the results of a Monte Carlo model. Red solid line: two body model, blue dashed line: three body model.  Bottom panels: Mandel Q parameter for each dataset.  For $n=41-43$, the horizontal lines represent the average Q value.  For $n=44$ and 45, the horizonal lines represent the average Q value for low and high densities, respectively.}
\end{figure}
\end{center}
\twocolumngrid

We excite atoms to $nD_{5/2}$ states, where $41 \leq n \leq 45$, for a range of ground state atom densities.  The top panels of Fig.~\ref{Fig2} show the slopes of the lines fit to the sorted data as a function of ground state atom density.  The slopes for $n=41$ and 42 are an approximately constant function of density at $\sim 2$, while for $n=43$ the slopes are $\sim 1.25$.  For $n=44$, the slopes are 1.6 for high density and get as large as 1.9 for low density, while for $n=45$ the slopes vary continuously from 1.55 at high density to 2.2 for low density.

The behavior of these graphs is correlated with the onset of the blockade as the density and principal quantum number are varied.  The data in the lower panels of Fig.~\ref{Fig2} represent the Mandel Q parameter for the atom number distributions represented in the upper panel.  The Mandel Q parameter is defined as $Q=\sigma^2 / \bar{N} -1$, where $\sigma^2$ is the dispersion in the total atom number and $\bar{N}$ is the mean number of counts.  The smaller $Q$, the narrower the distribution of excitation number, and the stronger the blockade~\cite{Cubel,Reinhard3}.  We can see that a stronger blockade is accompanied by a lower value for the slopes.  For $n=41$ and 42, the interactions are weak~\cite{Reinhard} and the system is not blockaded.  For $n=43$, the number of atoms per excitation domain ranges from 14 at the highest density to 2 at the lowest density, and the system is deeply in the blockaded regime.  For $n=44$ and 45, the number of atoms per excitation domain drops to less than 3 at approximately 7 and $6 \times 10^{10}$~cm$^{-3}$, respectively.  Thus for high densities, the system is blockaded with 4-5 atoms per excitation domain, while at low densities the excitation is uncorrelated~\cite{ARC}.

To understand our results, we consider the values the slopes of the sorted graphs would have, under ideal conditions.  Each excitation event may result in either a single atom in the target state ($nD_{5/2}$) or a state-mixing event.  In the blockaded regime, each mixing event will happen \emph{instead of} a single excitation to the target state in a single excitation domain.  In the regime of uncorrelated excitation, each mixing event will happen \emph{independently} of any other excitations.  Therefore, if state mixing occurs via an $m$-body interaction, each mixing event will add $m-1$ extra excitations in the blockaded regime and $m$ extra excitations in the unblockaded regime.  Thus, we expect slopes of 1 - 2 for two-body interactions, and slopes of 2 - 3 for three body interactions, depending on whether the system is blockaded or not.  All the slopes in Fig.~\ref{Fig2} fall in this range.

However, two factors make it difficult to unambiguously interpret the slopes.  Our detector efficiency is not one, and the number of excitation events fluctuates from shot to shot due to variations in ground state atom density, beam overlap, etc.  The latter effect causes the measured slopes to increase, since shots with more excitation events will be correlated with more atoms in $(n+2)P$, independent of our model for state mixing.  The former effect causes the measured slopes decrease since a lower detector efficiency causes some excitations to be missed, lessening the impact of the number fluctuations.

To account for these two effects, we have implemented a simple Monte Carlo model.  For each of $10^5$ iterations, we randomly draw the number of excitation events from a Gaussian distribution with a mean of 20.  We choose the width so that the final simulated distribution has the same $Q$ value (to within 0.005) as the data in the lower panels of Fig.~\ref{Fig2}.  For $n=41-43$, we use the average values indicated by the black horizontal lines.  For $n=44$ and 45, the $Q$ values vary between low and high density, so we use two atom number distributions with different widths.  We next randomly assign each excitation event to either a single atom in $nD_{5/2}$ or to a state-mixed event, using the experimentally measured mixing fractions (Fig.~\ref{Fig4}b).  For a two body process, a state-mixed event puts one atom each in $(n-2)F_{7/2}$ and $(n+2)P_{3/2}$.  For a three body process, we put one atom each in $(n-2)F_{7/2}$ and $(n+2)P_{3/2}$ and a third, spectator, atom in $nD_{5/2}$~\cite{Pohl}.  After simulating each excitation event, we randomly decide whether each Rydberg count is recorded, using an assumed detector efficiency.  Finally, we plot the total number of Rydberg counts vs. the number in $(n+2)P$ and fit a line to get the slope of the sorted data, just as in the experiment.

To generate predictions, we need to input a value for our unknown detector efficiency.  If we assume 3 body interactions, the slopes predicted by the model are close to the data for all $n$, except for $n=43$, using detector efficiencies in the range of $50-85\%$.  For $n=43$, a two-body model works better.  We use our detector efficiency as a free parameter to minimize the $\chi^2$ deviation between predicted and measured slopes for all $n$, simultaneously.  The best fit is $55\%$, which falls within the expected range~\cite{Wiza}.  The horizontal lines in the upper panel of Fig.~\ref{Fig2} are the results of our model for the assumption of two-body (red solid lines) and three-body (blue dashed lines) interactions. For our range of densities, state mixing is consistent with a three-body process for all $n$ except 43, while at $n=43$ a two-body process is a better fit~\cite{note}.  However, we cannot rule out a combination of two- and three-body interactions at $n=43$.

\begin{figure}[htp]
\centerline{ \scalebox{0.175} {\includegraphics{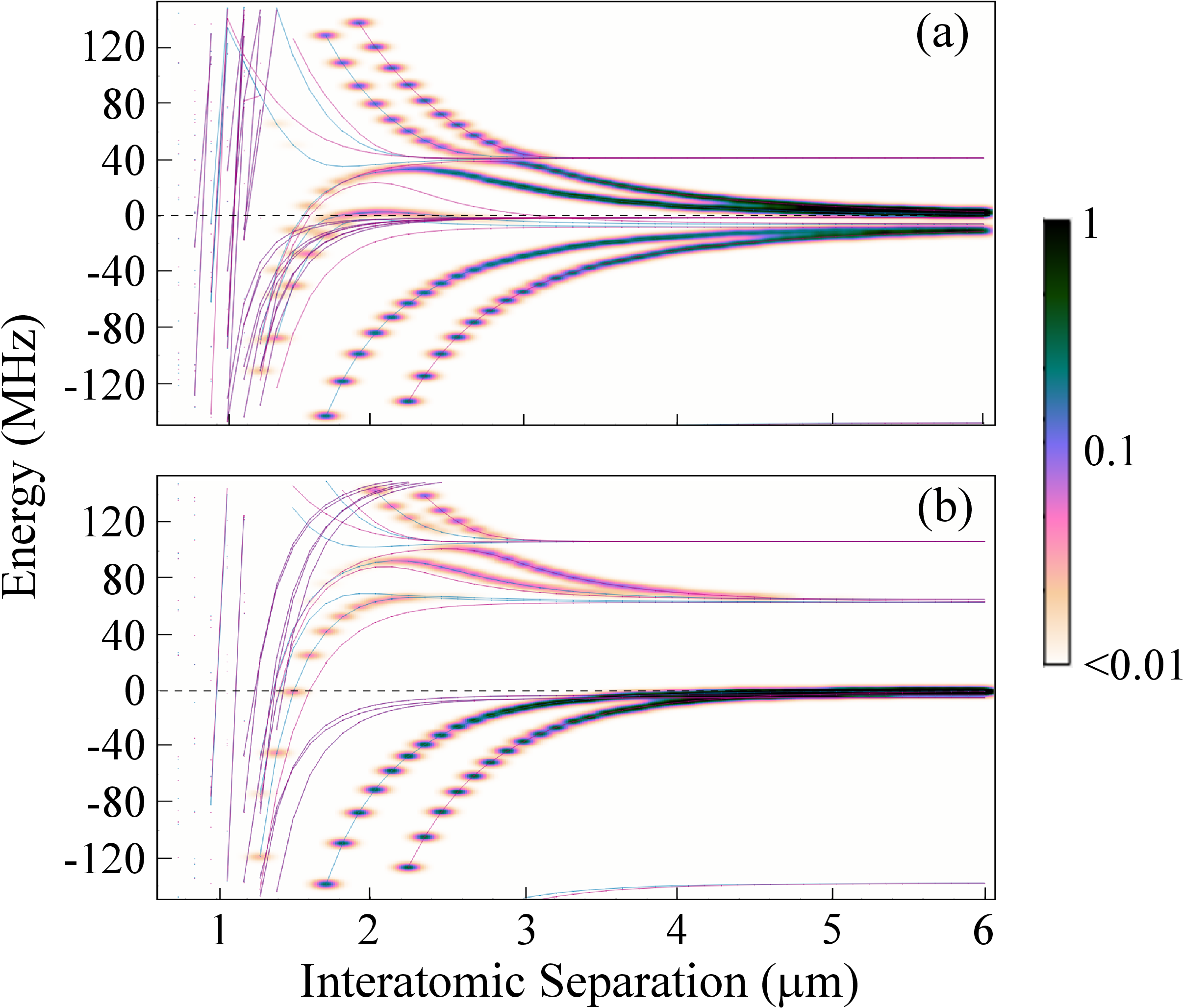}}} \caption{ \label{Fig3} Pair potential curves for two atoms $\vert nD_{5/2}, m_j=3/2 \rangle + \vert nD_{5/2}, m_j=5/2 \rangle$ state for $n=43$ (a) and $n=44$ (b). The angle between the atomic dipoles is zero. The weight of each curve indicates the overlap with the unperturbed state.}
\end{figure}

Since the large amounts of state mixing for $n=43$ are consistent with two-body interactions, it is important to identify a mechanism.  In Fig.~\ref{Fig3} we show calculations of the pair potential for two Rydberg atoms, including terms up to fifth order in the interaction operator~\cite{Weber,PairInteractionWebsite}.  We consider the states $\vert nD_{5/2}, m_j=3/2 \rangle + \vert nD_{5/2}, m_j=5/2 \rangle$, where $n=43$ (a) and $n=44$ (b).  In panel a, there are a range of distances between 1.5~$\mu$m and 3~$\mu$m for which a two particle potential asymptotically connected to $41F_{7/2} + 45P_{3/2}$ is near zero energy shift, while maintaining nonzero overlap with the target state.  Therefore, a range of pair distances could lead to excitation of this state.  In panel b, where the interaction channel is off-resonant, the zero crossing occurs at a single small radius, and with smaller overlap with the target state.  Therefore, this state is less likely to be excited.  A comprehensive study of the pair potentials would include a range of $n$, $m_j$, and angles.  However, our example illustrates the qualitative difference in the pair potentials when the dominant interaction channel is nearly resonant.

\begin{figure}[htp]
\centerline{ \scalebox{0.442} {\includegraphics{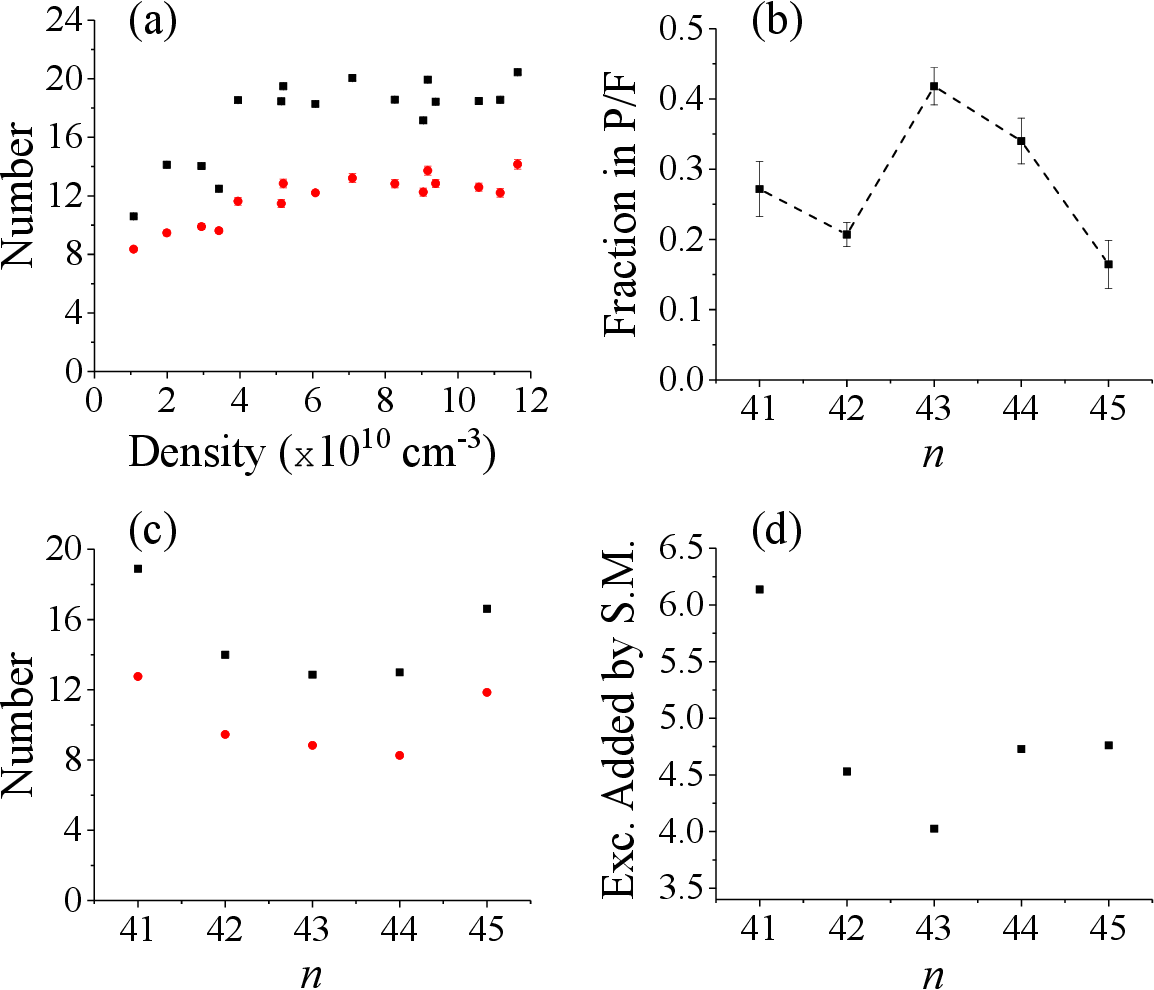}}} \caption{ \label{Fig4} (a) and (c): Total number of excitations (black squares) and intercepts of the sorted graphs (red circles) , as a function of density for $41D_{5/2}$ and as a function of $n$ at high density, respectively.  (b) Fraction of the averaged SSFI signal in $(n-2)F_{7/2}$ and $(n+2)P_{3/2}$ states as a function of principal quantum number at high density.  (d) The average number of excitations ``added'' by state mixing events.}
\end{figure}

While the slopes of the sorted graphs tell us about the mechanism underlying state mixing, the intercepts quantify the extent to which state mixing ``breaks the blockade.''  In Fig.~\ref{Fig4}a we plot the average number of excitations and the intercepts of the fits to the sorted data as a function of ground state atom density, for excitation to $41D_{5/2}$ states.  The asymptotic values of these curves are plotted as a function of $n$ in panel c. Both the total number of excitations and the y-intercept have a minimum for $n=43$ and 44, where the dominant interaction channel is approximately resonant.  In panel d, we plot the difference between the average number of excitations and the intercept of the sorted graphs, which is the number of excitations ``added'' by state mixing.  We see the counter-intuitive result that the \emph{number} of extra excitations added by state mixing is smallest for excitation to $n=43$, while the \emph{fraction} of atoms mixed into product states is largest (see Fig.~\ref{Fig4}b).  We attribute this to the strong Rydberg blockade for $n=43$.  The increased probability for a state-mixing event in a given excitation domain is overcome by the fact that, for stronger interactions, there are fewer total domains.  This provides evidence that the ``domain'' picture remains valid in the presence of state mixing.  Note also that the number of atoms created by state mixing is maximum for $n=41$, where interactions are weakest~\cite{Reinhard}.  The blockade picture breaks down most completely for this $n$, allowing for a larger number of excitation events.

In conclusion, we have demonstrated a new technique to quantify the impact of state mixing on the Rydberg excitation blockade.  By considering mixing on the single event level, we have achieved two goals: to determine how many extra excitations each mixing event adds, as well as the extent to which mixing breaks the blockade.  We conclude that, outside of quantum numbers with unfavorable molecular potential curves, large rates of state mixing near resonance are consistent with three-body interactions.  Therefore, systems with site-addressable single atoms will be immune to this decoherence mechanism, as long as a pair can be isolated. Future work will focus on looking for signatures of a specific three-body model~\cite{Pohl}.

{\em Acknowledgements.}  This work was supported by NSF Grant PHY-1553179.


\begin{thebibliography}{99}


\bibitem{Jaksch} D. Jaksch, J. I. Cirac, P. Zoller, S. L. Rolston, R. C\^ot\"e, and M. D. Lukin, Phys. Rev. Lett., \textbf{85} 2208 (2000)

\bibitem{Tong} D. Tong, S. M. Farooqi, J. Stanojevic, S. Krishnan, Y. P. Zhang, R. C\^ot\'e, E. E. Eyler, and P. L. Gould. Phys. Rev. Lett., \textbf{93} 063001 (2004)

\bibitem{Cubel} T. Cubel Liebisch, A. Reinhard, P. R. Berman, and G. Raithel, Phys. Rev. Lett. 95, 253002 (2005)

\bibitem{Urban} E. Urban, T. A. Johnson, T. Henage, L. Isenhower, D. D. Yavuz, T. G. Walker, and M. Saffman, Nature Physics \textbf{5}, 110 (2009)

\bibitem{Gaetan} Alpha Ga\"etan, Yevhen Miroshnychenko, Tatjana Wilk, Amodsen Chotia, Matthieu Viteau, Daniel Comparat, Pierre Pillet, Antoine Browaeys, and Philippe Grangier, Nature Physics \textbf{5}, 115 (2009)

\bibitem{Wilk} T. Wilk, A. Ga\"etan, C. Evellin, J. Wolters, Y. Miroshnychenko, P. Grangier, and A. Browaeys, Phys. Rev. Lett. \textbf{104}, 010502 (2010)

\bibitem{Dudin} Y. O. Dudin, L. Li, F. Bariani, and A. Kuzmich, Nature Physics \textbf{8}, 790 (2012)

\bibitem{Ebert} Matthew Ebert, Alexander Gill, Michael Gibbons, Xianli Zhang, Mark Saffman, and Thad G. Walker, Phys. Rev. Lett. \textbf{112}, 043602 (2014)

\bibitem{Labhun} Henning Labuhn, Daniel Barredo, Sylvain Ravets, Sylvain de L\'es\'eleuc, Tommaso Macr\'i, Thierry Lahaye, and Antoine Browaeys, Nature \textbf{534}, 667 (2016)

\bibitem{Reinhard_DR} A. Reinhard, K. C. Younge, T. Cubel Liebisch, B. Knuffman, P. R. Berman, and G. Raithel, Phys. Rev. Lett. \textbf{100}, 233201 (2008)

\bibitem{Reinhard3} A. Reinhard, K. C. Younge, and G. Raithel, Phys. Rev. A \textbf{78}, 060702(R) (2008)

\bibitem{Schwarzkopf} A. Schwarzkopf, R. E. Sapiro, and G. Raithel, Phys. Rev. Lett. \textbf{107}, 103001 (2011)

\bibitem{Schauss} Peter Schau\ss, Marc Cheneau, Manuel Endres, Takeshi Fukuhara, Sebastian Hild, Ahmed Omran, Thomas Pohl, Christian Gross, Stefan Kuhr, and Immanuel Bloch, Nature \textbf{491}, 87 (2012)

\bibitem{Li} L. Li, Y. O. Dudin, and A. Kuzmich, Nature, \textbf{498}, 466 (2013)

\bibitem{Barredo} D. Barredo, S. Ravets, H. Labuhn, L. B\'eguin, A. Vernier, F. Nogrette, T. Lahaye, and A. Browaeys, Phys. Rev. Lett., \textbf{112}, 183002 (2014)

\bibitem{Urvoy} A. Urvoy, F. Ripka, I. Lesanovsky, D. Booth, J. P. Shaffer, T. Pfau, and R. L\"ow, Phys. Rev. Lett. \textbf{114}, 203002 (2015)

\bibitem{Dudin2} Y. O. Dudin and A. Kuzmich, Science, \textbf{336} 887, (2012)

\bibitem{Baur} Simon Baur, Daniel Tiarks, Gerhard Rempe, and Stephan D\"urr. Phys. Rev. Lett., \textbf{112} 073901 (2014)

\bibitem{Tiarks} Daniel Tiarks, Simon Baur, Katharina Schneider, Stephan D\"urr, and Gerhard Rempe. Phys. Rev. Lett., \textbf{113}, 053602 (2014)

\bibitem{Pritchard} J. D. Pritchard, D. Maxwell, A. Gauguet, K. J. Weatherill, M. P. A. Jones, and C. S. Adams, Phys. Rev. Lett., \textbf{105}, 193603 (2010)

\bibitem{Hofmann} C. S. Hofmann, G. G\"unter, H. Schempp, M. Robert-de-Saint-Vincent, M. G\"arttner, J. Evers, S. Whitlock, and M. Weidem\"uller, Phys. Rev. Lett., \textbf{110}, 203601 (2013)

\bibitem{Maxwell} D. Maxwell, D. J. Szwer, D. Paredes-Barato, H. Busche, J. D. Pritchard, A. Gauguet, K. J. Weatherill, M. P. A. Jones, and C. S. Adams, Phys. Rev. Lett., \textbf{110}, 103001 (2013)

\bibitem{Altiere} Emily Altiere, Donald P. Fahey, Michael W. Noel, Rachel J. Smith, and Thomas J. Carroll, Phys. Rev. A, \textbf{84}, 053431 (2011)

\bibitem{Beguin} L. B\'eguin, A. Vernier, R. Chicireanu, T. Lahaye, and A. Browaeys, Phys. Rev. Lett., \textbf{110}, 263201 (2013)

\bibitem{Baluktsian} T. Baluktsian, R. L\"ow, H. K\"ubler, J. P. Shaffer, and T. Pfau, Nature Photonics, \textbf{4} 112, (2010)

\bibitem{Saffman_Rev} M. Saffman, T. G. Walker, and K. M{\o}lmer. Rev. Mod. Phys. \textbf{82} 2313, (2010)

\bibitem{Lukin} M. D. Lukin, M. Fleischhauer, R. C\^ot\"e, L. M. Duan, D. Jaksch, J. I. Cirac, and P. Zoller. Phys. Rev. Lett., \textbf{87} 037901(2001)

\bibitem{Isenhower} L. Isenhower, E. Urban, X. L. Zhang, A. T. Gill, T. Henage, T. A. Johnson, T. G. Walker, and M. Saffman, Phys. Rev. Lett. \textbf{104}, 010503 (2010)

\bibitem{Wang} Yang Wang, Xianli Zhang, Theodore A. Corcovilos, Aishwarya Kumar, and David S. Weiss Phys. Rev. Lett. \textbf{115}, 043003 (2015)

\bibitem{Wang2} Yang Wang, Aishwarya Kumar, Tsung-Yao Wu, David S. Weiss, Science \textbf{352}, 1562 (2016)

\bibitem{Weitenberg} Christof Weitenberg, Manuel Endres, Jacob F. Sherson, Marc Cheneau, Peter Schau\ss, Takeshi Fukuhara, Immanuel Bloch, and Stefan Kuhr, Nature \textbf{471}, 319 (2011)

\bibitem{Bernien} Hannes Bernien, Sylvain Schwartz, Alexander Keesling, Harry Levine, Ahmed Omran, Hannes Pichler, Soonwon Choi, Alexander S. Zibrov, Manuel Endres, Markus Greiner, Vladan Vuletic, and Mikhail D. Lukin, Nature \textbf{551}, 579 (2017)

\bibitem{Lienhard} Vincent Lienhard, Sylvain de L\'es\'eleuc, Daniel Barredo, Thierry Lahaye, Antoine Browaeys, Michael Schuler, Louis-Paul Henry, and Andreas M. L\"auchli, Phys. Rev. X 8, 021070 (2018)

\bibitem{deLesluec} Sylvain de L\'es\'eleuc, Sebastian Weber, Vincent Lienhard, Daniel Barredo, Hans Peter B\"uchler, Thierry Lahaye, and Antoine Browaeys, Phys. Rev. Lett. \textbf{120}, 113602 (2018)

\bibitem{Walker} Thad G. Walker and Mark Saffman, J. Phys. B \textbf{38}, S309-S319 (2005)

\bibitem{Walker2} Thad G. Walker and Mark Saffman, Phys. Rev. A, \textbf{77}, 032723 (2008)

\bibitem{Amthor} Thomas Amthor, Christian Giese, Christoph S. Hofmann, and Matthias Weidem\"uller, Phys Rev. Lett. \textbf{104}, 013001 (2010)

\bibitem{Reinhard} A. Reinhard, T. Cubel Liebisch, B. Knuffman, and G. Raithel, Phys. Rev. A, \textbf{75}, 032712 (2007)

\bibitem{Reinhard2} A. Reinhard, T. Cubel Liebisch, K. C. Younge, P. R. Berman, and G. Raithel, Phys. Rev. Lett. \textbf{100}, 123007 (2008)

\bibitem{Younge} K. C. Younge, A. Reinhard, T. Pohl, P. R. Berman, and G. Raithel, Phys. Rev. A \textbf{79}, 043420 (2009)

\bibitem{Pohl} T. Pohl and P. R. Berman, Phys. Rev. Lett. \textbf{102}, 013004 (2009)

\bibitem{Kondo} Jorge M. Kondo, Luis F. Goncalves, Jader S. Cabral, Jonathan Tallant, and Luis G. Marcassa, Phys. Rev. A \textbf{90}, 023413 (2014)

\bibitem{Orr} Keegan Orr, Ian George, and Aaron Reinhard, Rev. of Sci. Inst. \textbf{89}, 093107 (2018)

\bibitem{ARC} N. Sibalic, J. D. Pritchard, K. J. Weatherill, C. S. Adams, Computer Physics Communications \textbf{220}, 319 (2017)

\bibitem{Wiza} Joseph Ladislas Wiza, Nucl. Instrum. Methods \textbf{162}, 587 (1979).  Also posted at http://www.burle.com.

\bibitem{note} For $n=45$, the $nD_{5/2}$ peak runs into the $(n+2)P_{3/2}$ peak in the SSFI spectrum.  This causes the slopes of the sorted graphs at high density to be slightly lower than the three-body prediction.

\bibitem{Weber} S. Weber, C. Tresp, H. Menke, A. Urvoy, O. Firstenberg, H. P. B\"uchler, and S. Hofferberth, J. Phys. B \textbf{50}, 133001 (2017).

\bibitem{PairInteractionWebsite} https://pairinteraction.github.io.

\end{thebibliography}
\end{document}